\documentclass[useAMS,usenatbib,usegraphicx]{mn2e}
\usepackage{amsmath}
\def\simlt{\lower.5ex\hbox{$\; \buildrel < \over \sim \;$}}
\def\simgt{\lower.5ex\hbox{$\; \buildrel > \over \sim \;$}}

\def\beq{\begin{equation}}
\def\eeq{\end{equation}}
\def\rL{_{\rm L}}
\def\rn{_{\rm n}}
\def\rb{_{\rm b}}
\def\rf{_{\rm f}}
\def\rl{_{\rm l}}
\def\rs{_{\rm s}}
\def\rr{_{\rm r}}
\def\dc{\delta_{\rm c}}
\def\rtot{_{\rm tot}}
\def\rLSS{r_{\rm LSS}}
\def\Lmin{L_{\rm min}}
\def\Cmin{C_{\rm min}}
\def\Dmin{D_{\rm min}}

\title[Scale-Dependent Bias from BAOs]{Scale-Dependent Bias of
Galaxies from Baryonic Acoustic Oscillations} 

\author[Rennan Barkana and Abraham Loeb]{Rennan Barkana$^1$ and
Abraham Loeb$^2$\thanks{E-mail: barkana@wise.tau.ac.il (RB);
aloeb@cfa.harvard.edu (AL)}\\ $^1$ Raymond and Beverly Sackler School
of Physics and Astronomy, Tel Aviv University, Tel Aviv 69978,
Israel\\ $^2$Astronomy Department, Harvard University, 60 Garden
Street, Cambridge, MA 02138, USA}

\begin{document}

\pagerange{\pageref{firstpage}--\pageref{lastpage}} \pubyear{2010}
\maketitle
\label{firstpage}
\begin{abstract}
Baryonic acoustic oscillations (BAOs) modulate the density ratio of
baryons to dark matter across large regions of the Universe.  We show
that the associated variation in the mass-to-light ratio of galaxies
should generate an oscillatory, scale-dependent bias of galaxies
relative to the underlying distribution of dark matter. A measurement
of this effect would calibrate the dependence of the characteristic
mass-to-light ratio of galaxies on the baryon mass fraction in their
large scale environment. This bias, though, is unlikely to
significantly affect measurements of BAO peak positions.
\end{abstract}


\section{Introduction}\label{intro}

The rapid acoustic waves in the radiation-baryon fluid prior to
cosmological recombination were not followed by the dark matter at
that time. Following recombination, the baryons were freed from the
strong radiation pressure and fell into the gravitational potential
fluctuations of the dark matter. As a result, the fractional
difference between the density fluctuations of baryons and dark matter
decreased steadily with cosmic time. But since the baryons amount to a
sizable fraction of the total mass density of matter
($\Omega_b/\Omega_m\approx 17\%$), the gravitational effect of the
baryons on the dark matter imprinted baryonic acoustic oscillations
(BAOs) on the matter power spectrum. The characteristic comoving scale
of BAOs $\sim 100~{\rm Mpc}$ (corresponding to the sound horizon at
recombination), provides a yardstick that can be used to measure the
dependence of both the angular diameter distance and Hubble parameter
on redshift \citep[see review by][]{e05}.

When analyzing galaxy surveys, it is often assumed that galaxies are
biased tracers of the underlying matter distribution \citep{k84}, with
a bias factor that approaches a constant value on sufficiently large
scales where density fluctuations are still linear
\citep[e.g.,][]{mw96,tp98,smt01}.  However, the imprint of primordial
acoustic waves on the baryon fluid at recombination introduced a
scale-dependent modulation of the ratio between the density
fluctuations of baryons and dark matter that has not been completely
erased by the present time.  A large-scale region with a higher baryon
mass fraction than average (in the perturbations that lead to galactic
halos) is expected to produce more stars per unit total mass and hence
result in galaxies with a lower mass-to-light ratio.

In this paper we characterize the associated scale-dependent bias in
flux-limited surveys of galaxies.  The ratio between the power spectra
for fluctuations in the luminosity density and number density of
galaxies is expected to show BAO oscillations that reflect the
large-scale variations in the baryon-to-matter ratio.

In \S 2, we formulate the oscillatory BAO signature on galaxy bias in
terms of a simple analytical model. The quantitative results from this
model are presented in \S 3. Finally, we summarize our main
conclusions in \S 4.

\section{The Model}

\subsection{Basic Setup}

\label{basic}

Since galaxies sample the high peaks of the underlying matter density,
they are biased tracers of the matter density. When the clustering of
galaxies is usually analyzed, the bias is considered simply with
respect to the matter density, without separating out the effects of
the baryons. As long as the baryon fluctuations follow the same
spatial pattern as that of the dark matter, biasing with respect to
each of them cannot be separated since this separation is degenerate
with an overall change of the bias factor, which is not known
apriori. However, since the BAOs induce a scale-dependent difference
between the baryons and dark matter, it becomes important to consider
their influence on galaxies separately.

Consider the power spectrum of fluctuations in the galaxy number
density $n_{\rm gal}$ and in the luminosity density $\rho\rL$.
For a given galaxy population,
\beq \rho\rL = n_{\rm gal} \times \langle L \rangle\ , 
\eeq
where $\langle L \rangle$ is the mean luminosity of the
galaxies. Since galaxy formation is driven by halo collapse, which
depends on the evolution of the overall matter perturbations, the
number density fluctuations $\delta\rn$ are driven by the fluctuation
$\delta\rtot$ in the total matter density, with a bias $b\rn$ that
should be approximately constant on large scales (for a fixed galaxy
population):
\beq \delta\rn = b\rn \delta\rtot\ . \label{dn}\eeq
The mean luminosity of galaxies may depend on their environment
through their merger rate history, which is correlated with the local
matter density. This can lead to fluctuations $\delta\rL$ in
$\rho\rL$ with a different bias factor that should also approach
a constant on large scales:
\beq \delta\rL = (b\rn + b_{\rm L;t}) \delta\rtot\ , \label{eq:Lt} 
\eeq where the overall bias factor of the luminosity
density with respect to the total matter includes the number density
bias $b\rn$ as well as a possible additional bias $b_{\rm L;t}$ from
the dependence of $\langle L \rangle$ on the matter density.

However, the luminosity is also affected separately by the baryon
fluctuations, since the luminosity depends on the gas fraction in
halos $f\rb$. Regions that have halos with a higher baryon fraction
will proportionally have more baryons in the galaxies within them. If,
e.g., we assume that the star formation rate per baryon is on average
constant, then $\langle L \rangle\ \propto f\rb$. In fact, the
dependence of the luminosity on the gas fraction is likely to be
non-linear. For instance, in simple models for disk formation within
halos \citep{MMW}, the disk radius is approximately independent of the
gas fraction. Thus, if we assume that the disk mass is a fixed
fraction of the halo gas mass, then the typical gas surface density
within the disk varies in proportion to the overall halo gas
fraction. According to the Schmidt-Kennicutt law \citep[e.g.,][]{k98},
the star formation rate in the disk should vary with the gas surface
density to the power 1.4.  Thus, in general we assume that \beq
\langle L \rangle\ \propto (f\rb)^{b_{\rm L;f}}\ ,
\label{eq:L} \eeq
where these simple considerations suggest that $b_{\rm L;f}
\approx 1.4$. The notation for this power index is chosen since
equation~(\ref{eq:L}) (together with equation~\ref{eq:Lt}) implies
fluctuations
\beq \delta\rL = (b\rn + b_{\rm L;t}) \delta\rtot + 
b_{\rm L;f} \delta\rf\ , \label{dL} \eeq where $\delta\rf$ is the
perturbation in the halo gas fraction $f\rb$. Thus, $b_{\rm L;f}$ is
the bias factor of the luminosity density with respect to the halo
baryon fraction. Note that in our notation all the perturbations are
the actual ones at the considered redshift (i.e., we do not use the
common practice of linear extrapolation to redshift zero).

\subsection{Halo Baryon Fraction}

We would expect the baryon fraction within halos to reflect that of
their surroundings, but the precise relation is complex due to the
non-linear process of halo collapse. Here we employ reasonable
simplifications to derive an approximate analytical result, which is
partly verified and quantified by simulation results shown in \S~3.

We find it useful to analyze the baryon fraction in several steps,
where the first step is to avoid halo collapse and simply consider
\beq \gamma\rb \equiv \frac{\rho\rb} {\rho\rtot}\ , \eeq where we use
$\gamma\rb$ for the general baryon fraction and reserve $f\rb$ for the
baryon fraction inside halos. The mean of this quantity is the cosmic
mean baryon fraction: \beq \bar{\gamma}\rb =
\frac{\Omega_b}{\Omega_m}\ , \eeq and its fluctuation is simply \beq
\delta_\gamma = \delta\rb - \delta\rtot = r \delta\rtot\ .\eeq Here we
have measured the fractional difference between the baryonic and total
matter fluctuations with $r \equiv (\delta\rb/\delta\rtot)-1$, in
general a function of both wavenumber $k$ and redshift.

In reality, halos form out of perturbations that eventually grow to an
overdensity of hundreds, making the contribution of the mean density
negligible, and thus we expect the baryon fraction to reflect the
relative mass of the baryon perturbation that formed the halo:
\beq f\rb = \frac{ \Omega_b \delta\rb} { \Omega_m \delta\rtot} 
= \bar{\gamma}\rb \frac{ \delta\rb} { \delta\rtot}\ . \label{fb:1}
\eeq Before discussing non-linear collapse, we wish to apply this 
equation to the linear perturbations that will form a halo, but even
in the linear case we cannot easily apply this equation in Fourier
space, since halos form out of a sum of perturbations on all scales,
and taking a ratio as in equation~(\ref{fb:1}) is a non-linear
operation.

To make further progress, we make a separation of scales \citep[also
called a peak-background split;][]{Cole}, where we assume that the
fluctuations that we wish to observe (in the galaxy luminosity, etc.)
are on much larger scales than the (initial comoving) scales that
formed the halos. Typically, we are interested in measuring
fluctuations on BAO scales, which are $\sim 2$ orders of magnitude
above the halo formation scale of galaxies. Thus, we separate out the
linear halo perturbations (i.e., the initial perturbations that will
form a halo, linearly extrapolated to the formation redshift of a
particular halo):
\begin{align}
\delta\rtot &= \delta\rtot^l + \delta\rtot^s\ , \\
\delta\rb &= \delta\rb^l + \delta\rb^s = 
(1+r\rl) \delta\rtot^l + (1+r\rs) \delta\rtot^s\ ,
\end{align}
where the relative difference between the baryonic and total matter
perturbations is $r\rl$ and $r\rs$ on large and small scales,
respectively. 

We now use the standard result of spherical collapse, that a forming
halo has a linear $\delta\rtot = \dc$, where the critical density of
collapse $\dc$ is independent of mass (and equals 1.69 in the Einstein
de-Sitter limit, valid over a wide range of redshifts\footnote{The
value of $\dc$ decreases at low redshift due to the cosmological
constant, and at very high redshift due to the effects of the baryons
and radiation. However, at all $z<20$ the change is below $1\%$
\citep{NB07}.}). We also assume that we are considering sufficiently
large scales so that $\delta\rtot^l$ can be treated as a perturbation
of $\delta\rtot$ (or $\delta\rtot^s$), and that $r\rl$ and $r\rs$ are
also small quantities. Then the mean baryon fraction in halos is
\beq \bar{f}\rb = \bar{\gamma}\rb (1+r\rs)\ , \eeq
and the lowest order perturbation is derived to be
\beq \delta\rf = \frac{r\rl - r\rs}{\dc} \delta\rtot^l\ .
\eeq
We now use the actual value of $r(k)$ (see \S~3), specifically the
fact that it approaches a constant on scales below the BAOs, with a
value (depending on redshift but not $k$) that we denote $\rLSS$ (for
Large Scale Structure) following \citet{NB07}. Thus, in the
just-derived equations we can treat $r\rs = \rLSS$ as a constant (at a
given redshift), since most of the density $\dc$ needed to form a halo
comes from scales well below the BAO scale. Thus, the mean baryon
fraction in halos is
\beq \bar{f}\rb = \bar{\gamma}\rb (1+\rLSS)\ , \label{barf} \eeq 
while on large scales (i.e., small $k$) the fluctuation is
\beq \delta\rf = \frac{r(k) - \rLSS}{\dc} \delta\rtot\ .
\eeq

The remaining issue is the effect of non-linear collapse, and the
relation between the baryon fraction in the linearly-extrapolated halo
perturbation and the baryon fraction in the actual virialized halo.
We show simulation results in \S~3 that only test the mean baryon
fraction in halos (i.e., equation~\ref{barf}) but do so over a range
of redshifts, and suggest that halo collapse enhances the effect and
results in an effective value of $\rLSS$ that is amplified by a factor
of several. One way to understand this enhancement is to consider the
variation of $\rLSS$ with time. It declines (in absolute value)
approximately as $r \propto 1/a$ (where $a=1/(1+z)$ is the scale
factor), since $(\delta\rtot-\delta\rb) \approx $const while
$\delta\rtot \propto a$ (until the cosmological constant becomes
significant at low redshift). The decline of $\rLSS$ with time is of
critical importance, since we are computing it according to linear
theory, and it may not be appropriate to extrapolate $\rLSS$ all the
way to the halo formation time when we evaluate it in
equation~(\ref{barf}). The baryon fluctuations, which were erased on
small scales before cosmic recombination, later continuously catch up
with the dark matter (and thus with the total matter as well) in
linear perturbation theory. However, once a perturbation begins to
form a halo and enters the non-linear stage of collapse, we expect
that the rapid collapse will bring with it only the baryons already
present within the perturbation, and the continued decline of the
linear-theory $\rLSS$ will become irrelevant for the halo gas content.
The upshot is that the simulations suggest that if we use the
linear-theory $\rLSS$ (and, we assume, more generally for $r(k)$) then
we must multiply it by an effective amplification factor $A_r$:
\begin{align}
\bar{f}\rb &= \bar{\gamma}\rb (1+A_r\, \rLSS)\ ,\\
\delta\rf &= \frac{A_r}{\dc} [r(k) - \rLSS] \delta\rtot\ .
\end{align}
The resulting fluctuations in the luminosity density
(equation~\ref{dL}) are
\beq \delta\rL = (b\rn + b_{\rm L;t}) \delta\rtot + 
b_{\rm L;\Delta} [r(k) - \rLSS] \delta\rtot\ ,
\label{dL2} \eeq
where 
\beq b_{\rm L;\Delta} \equiv b_{\rm L;f} \frac{A_r}{\dc} \eeq
is an effective bias factor that measures the overall 
dependence of galaxy luminosity on the underlying difference
$\Delta$ between the baryon and total density fluctuations.

\subsection{Flux Limits}

We have assumed thus far that we observe a fixed galaxy population,
regardless of the varying luminosity of its members. In reality,
observed samples are limited by flux, or equivalently by luminosity if
for simplicity we consider galaxies at a single redshift. Suppose the
fraction of galaxies above luminosity $L$ is
\beq
F(L) = \int_{L^\prime = L}^{\infty} \phi(L^\prime) dL^\prime\ ,
\label{FL}
\eeq
where $\phi$ is the luminosity function. Then the observed number
density of galaxies is 
\beq n_{\rm obs} = n_{\rm gal} F(L)\ , \eeq
and the luminosity density of these galaxies is
\beq
\rho_L = n_{\rm gal} \langle L \rangle F(L)\ ,
\eeq
where 
\beq \langle L \rangle = \frac{1}{F(L)} \int_{L^\prime 
= L}^{\infty} L^\prime \phi(L^\prime) dL^\prime\ .
\eeq

We assume for simplicity that the same luminosity distribution holds
in different regions, except that the luminosity of all galaxies is
enhanced or diminished uniformly in response to changes in the total
density and the halo baryon fraction, as discussed in \S~\ref{basic}. If a
sample only includes galaxies above a detection threshold $\Lmin$,
then we can analyze the variations of $F(L)$ by keeping $\phi$ fixed
and varying the effective threshold $\Lmin$, while in $\rho\rL$ we
also include the perturbation in the luminosity of each galaxy. From
equation~(\ref{FL}) we obtain a relative fluctuation
\beq
\delta_F = \Cmin [b_{\rm L;t}
\delta\rtot + b_{\rm L;f} \delta\rf]\ ,
\eeq
where the dimensionless coefficient
\beq
\Cmin =
\frac{\Lmin\, \phi(\Lmin)}{F(\Lmin)}\ .
\eeq

The dependence of luminosity on the halo baryon fraction introduces a
dependence of the galaxy number density on the baryon fluctuations
(i.e., on $r(k)$). Putting our results together, for a flux-limited
survey we find
\beq \label{dn2}
\delta\rn = (b\rn + \Cmin b_{\rm L;t}) \delta\rtot + 
\Cmin b_{\rm L;\Delta} [r(k) - \rLSS] \delta\rtot\ ,
\eeq
and
\beq
\delta\rL = [b\rn + (1+\Dmin) b_{\rm L;t}] \delta\rtot + 
(1+\Dmin) b_{\rm L;\Delta} [r(k) - \rLSS] \delta\rtot\ ,
\eeq
where
\beq
\Dmin = \frac{\Lmin}{\langle L \rangle} \Cmin\ ,
\eeq
with $\langle L \rangle$ evaluated for $L=\Lmin$.

In the limit where $\Lmin$ is well below the peak of the luminosity
function, $\Cmin$ and $\Dmin$ both approach zero, and these
expressions simplify to the previous ones (equations~\ref{dn} and
\ref{dL2}). In the opposite limit, e.g., in the exponential
tail of a Schechter function, we can approximately set $\phi(L)
\propto e^{-L/L_*}$, and then $\Cmin = \Lmin / L_*$ and 
$\Dmin = \Cmin \Lmin/(\Lmin + L_*)$ are both $\gg 1$ when
$\Lmin \gg L_*$.

\subsection{Observational Proposals}

\label{s:goals}

As we have shown, both the galaxy luminosity density and (for a
flux-limited sample) number density depend on the halo gas fraction.
The scale-dependence of the relation between the baryon and dark
matter fluctuations implies that the BAOs can be observed in ratios
that previously would have been expected to be scale-independent.

One proposal is to compare the power spectrum of fluctuations in the
galaxy number density ($P\rn$) with that of the luminosity density
($P\rL$), with both measured for the same galaxy sample. Taking the
ratio may help to clear away some systematic effects that affect both
power spectra. Their ratio (square-rooted) should have the form
(assuming $r(k) \ll 1$):
\beq \left( \frac{P\rL} {P\rn} \right)^{1/2} = 
B_1 \left\{1 + B_2 [r(k) - \rLSS]\right\} \ , \label{Ps}
\eeq
where the various bias factors enter into the coefficients $B_1$ and
$B_2$. If we denote the bias ratio $b\rr \equiv b_{\rm L;t} / b\rn$,
then
\beq B_1 = \frac{1 + (1+\Dmin) b\rr} {1 + \Cmin b\rr}\ , \eeq
and 
\beq B_2 = \frac{b_{\rm L;\Delta}}{ b\rn} \frac{1 + 
\Dmin - \Cmin} {(1 + \Cmin b\rr) \cdot [1 + (1+\Dmin) b\rr]}\ .\eeq 
Note that in the limit where most of the galaxy population is observed
(i.e., the flux limits are unimportant), these expressions simplify to
$B_1 = 1 + b\rr$ and $B_2 = b_{\rm L;\Delta} /(b\rn B_1)$.

In practice, using these expressions is not as daunting as it may
appear. For a given galaxy sample, $\Cmin$ and $\Dmin$ can be
calculated from the measured luminosity function. This leaves two
unknowns, $b\rr$ and the ratio $b_{\rm L;\Delta} /b\rn$. Within the
ratio, we have a well-motivated expectation for $b_{\rm L;\Delta} =
b_{\rm L;f} A_r/\dc$, given that $\dc \approx 1.7$, $b_{\rm
L;f} \approx 1.4$ (\S~2.1), and $A_r \approx 3$ from simulations
(see \S~3). Now, if $r$ were independent of scale, then we
could only measure a degenerate combination of the unknown
quantities. However, a precise measurement of the power spectrum
ratio can separate out the constant and BAO terms, thus yielding
$B_1$ and $B_2$ separately, which in turn yields $b\rr$
and the ratio $b_{\rm L;\Delta} /b\rn$.

Although it is implicit in the equations, $r(k)$ and $\rLSS$ are also
declining functions of time. However, even at low redshift $r(k)$
contains a signature of the BAOs, since the BAOs are still imprinted
more strongly in the baryon fluctuations than in those of the dark
matter or the total matter. This clear signature offers a chance to
detect this effect, even if the various bias factors that we have
assumed to be constant actually vary slowly with $k$. A detection of
the effect can be combined with an estimate of $b\rn$ from comparing
$P\rn$ with the underlying matter power spectrum (e.g., as measured
with weak lensing on large scales). Extraction of the value of $b_{\rm
L;\Delta}$ would yield a new quantity in galaxy formation, a
combination of the way in which the luminosity of a galaxy depends on
the baryonic content of its host halo, and of how this baryonic
content depends on the underlying difference between the baryon and
total density fluctuations.

Another possibility is to compare the power spectra of luminosity
density (or flux-limited number density) between two different
samples. Their ratio should again have a form similar to
equation~(\ref{Ps}), from which the constant and BAO term can be
separately measured.  It is well known that galaxy bias depends on
galaxy luminosity \citep{l96}, but here the bias would be scale
dependent in a way that depends on $L_{\rm min}$.

\section{Quantitative Predictions}

For our quantitative results, we use the CAMB linear perturbation code
\citep{CAMB}, with the WMAP 5-year cosmological parameters \citep{WMAP5},
matching the simulation that we compare with below.

\begin{figure}
\includegraphics[width=84mm]{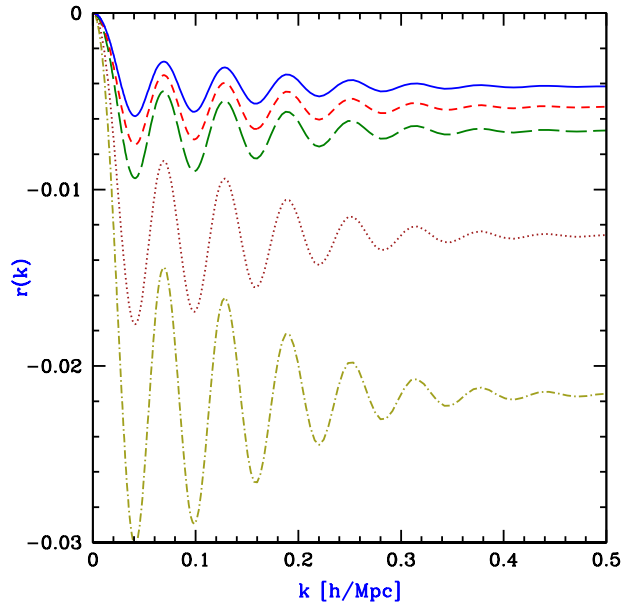}
\caption{The fractional baryon deviation 
$r(k)=(\delta\rb/\delta\rtot)-1$ as a function of $k$, at various
redshifts ($z=0$, 0.5, 1, 3, and 6, from top to bottom).}
\label{f:r}
\end{figure}

\begin{figure}
\includegraphics[width=84mm]{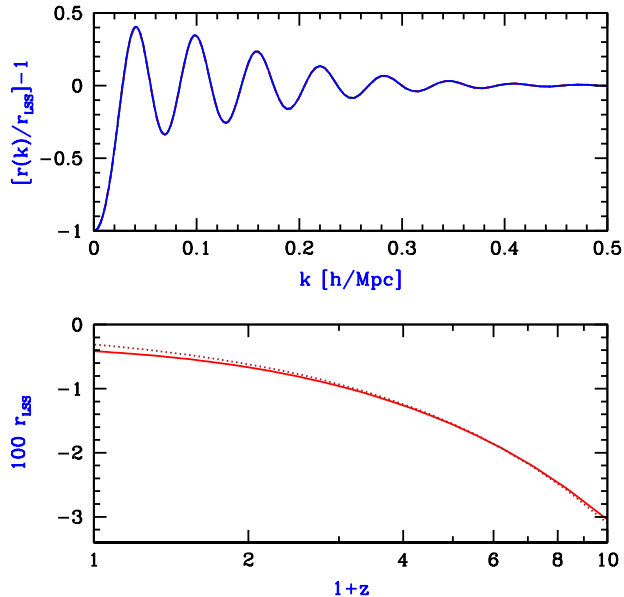}
\caption{Top panel: $[r(k)/\rLSS]-1$ as a function of $k$, 
at the same redshifts as in Figure~\ref{f:r} (the curves all lie on
top of each other). Bottom panel: The quantity $100 \rLSS$ versus
$1+z$ (solid curve), or equivalently, the value of $\rLSS$ in units of
percent. Also shown is the function $-0.31/a$ (dotted curve). For both
panels, in practice we set $\rLSS \equiv r(k=1{\rm h/Mpc})$.}
\label{f:r2}
\end{figure}

We show the dependence of $r$ on both wavenumber and redshift in
Figure~\ref{f:r}. At a given redshift, $r(k)$ approaches a constant
value ($\rLSS$) at $k \simgt 0.5$~h/Mpc. Using $\rLSS$ (itself a
function only of redshift) we can separate out the two variables $k$
and $z$ in their effect on $r$, as shown in Figure~\ref{f:r2}. The
function $[r(k)/\rLSS]-1$ is independent of redshift (i.e., the curves
for five different redshifts overlap very precisely), so the $k$
dependence of $r$ is determined by a single, fixed function of
$k$. Thus, the redshift dependence of $r$ is the same at all $k$, and
it suffices to show the dependence of $\rLSS$.  Figure~\ref{f:r2}
shows that, as noted in the previous section, $\rLSS$ indeed varies
approximately in proportion to $1/a$, but in detail the variation with
redshift is slightly slower than that.

As noted in the previous section, we expect the non-linear evolution
that takes place during halo formation to magnify the gas depletion
effect compared to the linear theory calculation. We can test this
effect using the hydrodynamical simulation of \citet{NYB}. Although
superficially it appears that they studied a quite different regime
(low-mass halos forming at high redshift), their results should be
applicable here. In the linear theory, the gas depletion factor
$\rLSS$ is constant all the way from nearly the BAO scale ($k \sim
0.5$~ h/Mpc) down to just above the Jeans scale ($k \simgt 100$~
h/Mpc). \citet{NYB} investigated the gas depletion in virialized halos
from below the Filtering mass (which is a time-averaged Jeans mass) up
to a thousand times higher mass scale. Thus, the most massive halos in
their simulation were well into the large-scale structure regime,
where pressure is negligible, and the effect we are interested in
(i.e., non-linear gas depletion on large scales) should operate.

\begin{figure}
\includegraphics[width=84mm]{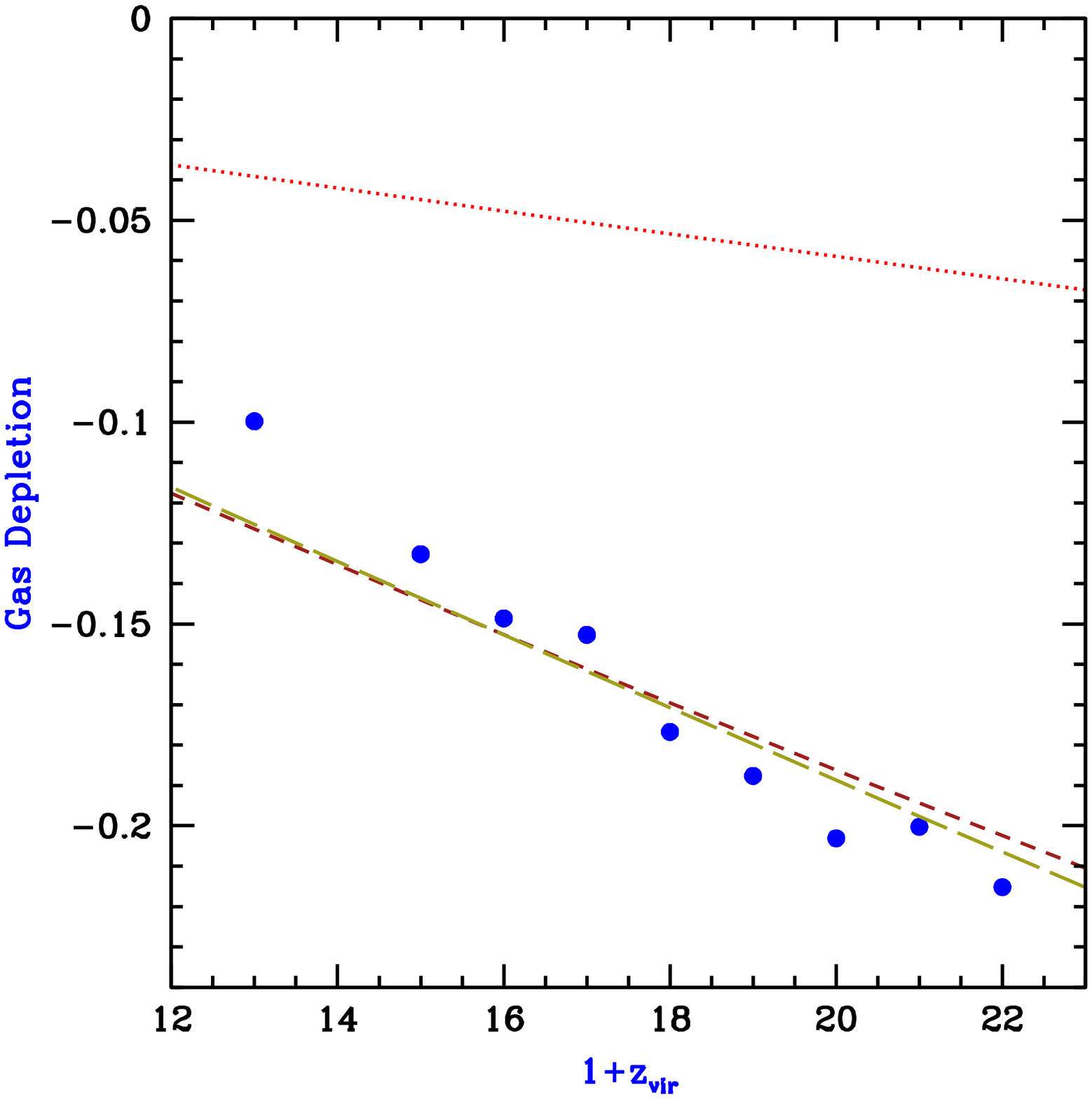}
\caption{The fractional gas depletion in halos versus redshift.
We show the results from the simulation of \citet{NYB} (data points),
where the corresponding redshift (at which the virialized halos are
identified in the simulation) is denoted $z_{\rm vir}$. We compare the
depletion as measured in the simulation to $\rLSS$ at $z_{\rm vir}$
(dotted curve), $3.2 \rLSS$ at $z_{\rm vir}$ (long-dashed curve), and
$\rLSS$ at a value of $(1+z)$ equal to $3.5 (1+z_{\rm vir})$
(short-dashed curve).}
\label{f:NYB}
\end{figure}

Figure~\ref{f:NYB} shows that the fractional gas depletion measured in
virialized halos in the simulation of \citet{NYB} was much larger than
the depletion $\rLSS$ predicted for linear perturbations at the halo
virialization redshift $z_{\rm vir}$. The simulated results can be
reasonably fit\footnote{There is a hint of a different slope with
redshift in the simulation results compared to the fits. However, this
mainly depends on a single point (at $z_{\rm vir}=12$) and needs to be
checked with further simulations.} either by multiplying $\rLSS$ by a
factor of 3.2, or by adopting $\rLSS$ from a higher redshift $z$
[where $(1+z) = 3.5 (1+z_{\rm vir})$]. Additional simulations are
required to test whether these results can indeed be extrapolated to
our regime of much more massive halos at low redshift, but these
results suggest that the gas depletion in halos is amplified by a
factor $\simgt 3$ compared to the linear regime.

As an example of typical numbers, we consider an example with
$\Cmin=\Dmin=0$, $b\rn=2$, and $b_{\rm L;t}=1$. As noted in
\S~\ref{s:goals}, we expect in this case $b_{\rm L;\Delta} \sim
2.6$, and also $b\rr=0.5$, so in the observational ratio of
equation~(\ref{Ps}), $B_1 = 1.5$ and $B_2 = 0.9$. Thus, the
oscillations in the square-rooted ratio of the luminosity and number
density power spectra are at the level of $0.4\%$ at $z=1$ (measured
from the first peak, i.e. at the lowest $k$, to the following trough;
the variation from $k=0$ to the first peak is roughly twice as
large). This is a weaker effect by about a factor of five compared to
the normal BAOs in the total matter power spectrum. Thus, if high
precision is achieved in the regular BAO measurement, then the
scale-dependent bias that we have highlighted should also be
measurable.

This scale-dependent bias is unlikely to significantly affect the
standard BAO measurements. Such measurements are usually carried out
on the power spectrum of the galaxy number density. Scale-dependent
bias enters this quantity only in proportion to $\Cmin$ (see
equation~\ref{dn2}), so it would be present only in a sample for which
the flux limit plays a significant role. Even then, the effect on the
BAO peak positions would be quite weak, since the BAOs in
$\delta\rtot$ are physically a result of the influence of the baryons
on the dark matter. Thus, the peak positions in $\delta\rtot$ and in
$\delta_b$ are nearly identical. For instance, even in the case that
in equation~(\ref{dn2}) the coefficients $(b\rn + \Cmin b_{\rm L;t})$
and $\Cmin b_{\rm L;\Delta}$ are equal, the BAO peak positions are
shifted by only $\sim 0.3\%$.

\section{Conclusions}

We have shown that the variation in the baryon to matter ratio
imprinted by acoustic waves prior to cosmological recombination should
result today in an oscillatory, scale-dependent bias of galaxies
relative to the underlying matter distribution (see Figs.~1 \& 2).
The percent-level amplitude of this signature depends on how the
typical luminosity of galaxies scales with the baryon mass fraction in
the large-scale region in which they reside. Simulations suggest that
this signature is significantly amplified by non-linear effects during
halo collapse (Fig.~3). The resulting amplitude can be measured from
the ratio between the power spectra of fluctuations in the luminosity
density and number density of galaxies (equation~\ref{Ps}). An
observational calibration of this amplitude would offer a new
cosmological probe of the physics of galaxy formation.

This effect may be marginally observable with current data, but it
should certainly be feasible using future galaxy surveys (such as
BOSS\footnote{http://cosmology.lbl.gov/BOSS/} or
BigBOSS\footnote{http://bigboss.lbl.gov/index.html}). However, since
the baryonic and the matter fluctuations have nearly identical BAO
peak positions, the scale-dependent bias is unlikely to significantly
affect the standard BAO measurements, even at percent-level precision.

\section*{Acknowledgments}
We thank the US-Israel Binational Science Foundation for grant support
that enabled this collaboration (BSF grant 2004386). This work was
also supported in part by Israel Science Foundation grant 823/09 (for
R.B.), and NSF grant AST-0907890 and NASA grants NNX08AL43G and
NNA09DB30A (for A.L.).


\label{lastpage}

\end{document}